# Nucleon masses in strong magnetic fields

H.R. Rubinstein[a] S. Solomon[b] and T. Wittlich[c]

[a]Department of Radiation Sciences, University of Uppsala, Uppsala, Sweden

[b]Racah Institute of Physics, Hebrew University of Jerusalem, Givat Ram, Jerusalem, Israel

[c]Physikalisches Institut, Universität Bonn, Nußallee 12, 53115 Bonn, Germany

We compute the variation of the masses of the proton and the neutron induced by the presence of a strong external magnetic field. We discuss the choice of the wave function and different techniques how to apply the magnetic field. The results obtained from a $24 \cdot 16^2 \cdot 12$ and a $16^2 \cdot 8^2$ lattice are compared with a phenomenological approach.

## 1. INTRODUCTION

We study the variation of the masses of the proton and the neutron induced by the presence of a strong external magnetic field. Our work is inspired by recent results of M.Bander and H.R.Rubinstein [1]. They consider the behavior of proton, neutron and electron in the presence of a constant magnetic field $B$.

Their treatment assumes that for sufficiently small magnetic field the dependency of the nucleon masses on $B$ is controlled by their anomalous magnetic moments which are known from experiment. The masses of proton and neutron can easily be computed using Landau's formula for a Dirac particle in a constant magnetic field [2]:

$$E_s^p = m_p + \frac{eB}{2m_p}(1 - g_p s) \qquad (1)$$

$$E_s^n = m_n + \frac{eB}{2m_n} g_n s \qquad (2)$$

$s = \pm 1$ denotes the spin, $m_p$ and $m_n$ the nucleon masses for $B = 0$. The formulae hold for zero momentum, the lowest Landau level and $eB > 0$.

Contributions to the masses which are even in the magnetic field such as $|eB|$ and $|eB|^2$ are assumed absent in (2). Since such terms are not measurable by usual Zeeman experiments and cancel in the g-factor, the only way to estimate them is either experimentally (i.e. observing the conditions described in [1]) or by lattice simulations. Usually the presence of non-analytic terms of the type $|eB|$ is discarded in phenomenological formulae on physical grounds. However, such non-analytic terms do appear in the formulae for the point Dirac particle [2] and might survive in the low energy hadron masses too.

Our simulations reported below are designed to put limits on these contributions. In fact, we confirm within the precision limits of our statistics the implicit hypothesis in [1] that such terms are absent.

Ref [1] pointed out that if one inserts the experimentally known anomalous magnetic moments, $g_n = -1.9$ and $g_p = 2.79$, then for increasing $B$ the lowest state of the neutron goes down steeper than the lowest state of the proton. Consequently, since $m_n > m_p$ there is a value of the magnetic field $B_0$ where the masses of the two particles are equal. $B_0$ is a very large field of the order of $10^{14}T$. Ref [1] also pointed out that for even larger fields ($> 10^{16}T$) a $\beta$–decay of the proton may become possible. Fields of that order of magnitude may appear in certain astronomical phenomena. An unstable proton is certainly an interesting effect by itself. It also has implications for astrophysics (see [1] for references).

However, the magnetic field required is very large, so that the assumption of pointlike particles does not necessarily hold and the effect of the strong forces has to be taken into account. Since there is no analytic nonperturbative approach in QCD, it is worthwhile to treat the problem by means of lattice QCD.

External magnetic fields have already been used as a tool to compute g–factors of proton and neutron [4]. In the present paper we study the spin up/down states of proton and neutron and compare them which the phenomenological approach of [1]. In particular we compute the slopes ($energy/B$) which is in a sense a harder task than computing the g–factors. For the g–factors it is sufficient to measure the mass difference of spin up and down states which is an observable that is more stable against systematic and statistical errors. In particular it is insensitive to terms even in the magnetic field, and to the QCD contributions to the overall hadronic mass.

## 2. THE METHOD

We use a method similar to the one described in [4]. We work with a set of 30 quenched $SU(3)$ configurations separated by 500 sweeps with 2000 sweeps thermalisation, $\beta = 6.0$. We use lattice sizes of $16^2 \cdot 8^2$ and $24 \cdot 16^2 \cdot 12$ and Wilson fermions at $K = 0.1475$ resp. $K = 0.145$.

On the lattice a uniform magnetic field $B$ can be introduced by multiplying a link $U_\mu(\tilde{x})$ by a phase $U_\mu^B(\tilde{x})$. For a field in $z$ direction we set:

$$U_x^B(\tilde{x}) = \exp(-ieBa^2 yL_x) \quad \text{for } x = L_x - 1$$
$$= 1 \quad \text{otherwise}$$
$$U_y^B(\tilde{x}) = \exp(ieBa^2(x - x_0))$$

where $x_0$ is an offset for the magnetic field to be discussed below.

Consequently the plaquette in the x–y plane is

$$P^B(\tilde{x}) = \exp(ieBa^2(-L_x L_y + 1))$$
$$\text{for } x = L_x - 1, y = L_y - 1$$
$$= \exp(ieBa^2) \quad \text{otherwise}$$

Thus the magnetic flux is homogeneous only if $B$ is quantized: $a^2 eB = 2\pi n/L_x L_y$.

Unfortunately the lowest possible $B$ is very large for reasonable sizes of the lattice. Due to the limited computer power available to us we decided to ignore the quantization. This introduces an inhomogenuity at $x = L_x - 1, y = L_y - 1$. In order to limit its influence we put the source in the middle of the lattice and increased the lattice size in the $x - y$–plane.

We found that the masses depend on the choice of the offset $x_0$ in relation to the position of the source $x_s$. In order to get control about the parameters we compared the 'flat' case (all $SU(3)$ links set to one) with the predictions from the continuum. In particular we computed the single $SU(6)$ 'proton' and 'neutron' states and compared their dependence on $B$ with the prediction given by the Landau formula [2] applied separately to each quark in the $SU(6)$ wave function. We expect that the proton spin up and the neutron spin down states should not depend on $B$ while the states with opposite spin should go up for increasing $B$.

If we compute these states for different choices for the offset $x_0$ we find a drastic difference between $x_0 = x_s \equiv L_x/2$ and $x_0 = 0$.

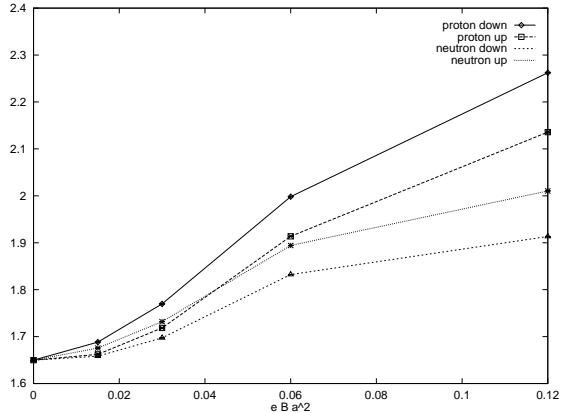

Figure 1. no QCD, $16^2 \cdot 8^2$: effective masses for $x_0 = 0$. $\diamond = p_\downarrow, \square = p_\uparrow, \triangle = n_\downarrow, \times = n_\uparrow$.

Figs. 1 and 2 show the effective masses at $t = 5$ (which is representative for all $t$). While the choice $x_0 = x_s$ reproduces the expected result the choice $x_0 = 0$ shows both proton states becoming heavier than the neutron states for sufficiently large $B$. This 'effect' can also be seen qualitatively when QCD is switched on and may be confounded with the physical effect predicted in [1].

In order to extract hadron masses from the correlation function the wave function of the particle of interest should have maximal overlap with the physical state. One way to achieve a better overlap is certainly to use non-pointlike sources. Due to the interaction of the magnetic field configuration with the position of the source we have considered the simplest case of pointlike sources.

|  | phenom | $V = \infty$ | $24 \cdot 16^2 \cdot 12$ | $16^2 \cdot 8^2$ |
|---|---|---|---|---|
| neutron ↓ slope | −1.9 | −2.1 | −2.1(4) | −2.2(4) |
| neutron ↑ slope | +1.9 | +1.3 | +1.9(3) | +3.2(2) |
| neutron g–factor $g_n$ | −1.9 | −1.7 | −2.0(4) | −2.6(4) |
| proton ↓ slope | +3.8 | +3.5 | +3.4(4) | +3.3(4) |
| proton ↑ slope | −1.8 | −1.9 | −2.9(4) | −4.8(2) |
| proton g–factor $g_p$ | +2.8 | +2.6 | +3.1(4) | +4.1(5) |
| $g_p/g_n$ | −1.47 | −1.52 | −1.52(3) | −1.56(4) |

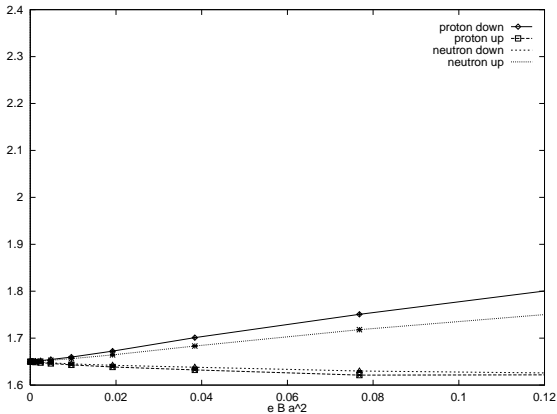

Figure 2. no QCD, $16^2 \cdot 8^2$: effective masses for $x_0 = x_s$. $\diamond = p_\downarrow, \square = p_\uparrow, \triangle = n_\downarrow, \times = n_\uparrow$.

We find that the overlap is much more sensitive to the choice of the wave function than it is for 'non–magnetic' mass simulations. In fact we find the $SU(6)$ wave function to have a better overlap than the $(u\gamma_5 Cu)d$ wave function [3].

## 3. RESULTS

We computed the slopes $(m(B) - m(0))\frac{2m(B)}{eB}$, the g–factors for proton and neutron and the ratio of their magnetic moments. The table shows the numerical results in comparison with the experimental data. The data stem from a $16^2 \cdot 8^2$ lattice at $K = 0.1475$, $a^2 eB = 0.03$, and from a $24 \cdot 16^2 \cdot 12$ lattice at $K = 0.145$, $a^2 eB = 0.02$. They differ for the two lattice sizes, in particular for the spin up states. If we extrapolate to infinite volume by assuming a $a = a_0 + 1/V a_1$ behavior we get results which are remarkably close to the experimental data.

The ratios of the g–factors have been extracted from the expression [4]

$$\frac{g_p}{g_n}(t) = \frac{\{\frac{G_{p\uparrow}-G_{p\downarrow}}{G_{p\uparrow}+G_{p\downarrow}}\}_t - \{\frac{G_{p\uparrow}-G_{p\downarrow}}{G_{p\uparrow}+G_{p\downarrow}}\}_{t-1}}{\text{same but } p \leftrightarrow n} \quad (3)$$

Since this expression contains only ratios of masses we expect smaller systematical errors from the pointlike sources. Indeed $\frac{g_p}{g_n}(t)$ is constant within 2% for all timeslices $t \geq 3$. These results also hold for $a^2 eB = 0.06$ resp. $0.08$.

Our data for the slopes lead to the following conclusions: We find a dependency on the volume $V$. However, in the limit of infinite $V$ the slopes give strong evidence that the linear phenomenological approach of ref [1] holds in the strong field regime.

In order to decide whether the proton becomes heavier than the neutron more precision is necessary. Further investigations thus should include:
- a detailed study of the finite size effects
- better statistics
- smeared sources adapted to the problem
- extrapolation to the chiral limit.